\documentclass[twocolumn,tighten,times]{aastex631}

\usepackage{microtype}
\usepackage{xspace}
\usepackage{amsmath}
\usepackage{textcomp}

\newcommand{\Msun}{\ensuremath{{\rm M}_{\sun}}\xspace}

\newcommand{\artis}{\textsc{artis}\xspace}

\shorttitle{3D radiative transfer for kilonovae}
\shortauthors{Shingles et al.}

\begin{document}

\title{Self-consistent 3D radiative transfer for kilonovae: directional spectra from merger simulations}

\correspondingauthor{Luke Shingles}
\email{luke.shingles@gmail.com}

\author[0000-0002-5738-1612]{Luke J. Shingles}
\affiliation{GSI Helmholtzzentrum für Schwerionenforschung,
Planckstraße 1, 64291 Darmstadt, Germany}

\author[0000-0002-0313-7817]{Christine E. Collins}
\affiliation{GSI Helmholtzzentrum für Schwerionenforschung,
Planckstraße 1, 64291 Darmstadt, Germany}

\author[0000-0002-4690-2515]{Vimal Vijayan}
\affiliation{GSI Helmholtzzentrum für Schwerionenforschung,
Planckstraße 1, 64291 Darmstadt, Germany}
\affiliation{Department of Physics and Astronomy, Ruprecht-Karls-Universität Heidelberg, Im Neuenheimer feld 226, 69120 Heidelberg, Germany}

\author[0000-0003-2024-2819]{Andreas Flörs}
\affiliation{GSI Helmholtzzentrum für Schwerionenforschung,
Planckstraße 1, 64291 Darmstadt, Germany}

\author[0000-0002-3126-9913]{Oliver Just}
\affiliation{GSI Helmholtzzentrum für Schwerionenforschung,
Planckstraße 1, 64291 Darmstadt, Germany}
\affiliation{Astrophysical Big Bang Laboratory, RIKEN Cluster for Pioneering Research, 2-1 Hirosawa, Wako, Saitama 351-0198, Japan}

\author[0000-0002-0093-0211]{Gerrit Leck}
\affiliation{GSI Helmholtzzentrum für Schwerionenforschung,
Planckstraße 1, 64291 Darmstadt, Germany}
\affiliation{Institut {f\"ur} Kernphysik (Theoriezentrum), Technische Universit{\"a}t Darmstadt, Schlossgartenstra{\ss}e 2, D-64289 Darmstadt, Germany}

\author[0000-0002-2385-6771]{Zewei Xiong}
\affiliation{GSI Helmholtzzentrum für Schwerionenforschung,
Planckstraße 1, 64291 Darmstadt, Germany}

\author[0000-0001-6798-3572]{Andreas Bauswein}
\affiliation{GSI Helmholtzzentrum für Schwerionenforschung,
Planckstraße 1, 64291 Darmstadt, Germany}

\author[0000-0002-3825-0131]{Gabriel Mart\'inez-Pinedo}
\affiliation{GSI Helmholtzzentrum für Schwerionenforschung,
Planckstraße 1, 64291 Darmstadt, Germany}
\affiliation{Institut {f\"ur} Kernphysik (Theoriezentrum), Technische Universit{\"a}t Darmstadt, Schlossgartenstra{\ss}e 2, D-64289 Darmstadt, Germany}

\author[0000-0002-9774-1192]{Stuart A. Sim}
\affiliation{Astrophysics Research Centre, School of Mathematics and Physics, Queens University Belfast, Belfast BT7 1NN, UK}




\begin{abstract}
We present three-dimensional radiative transfer calculations for the ejecta from a neutron star merger that include line-by-line opacities for tens of millions of bound-bound transitions, composition from an r-process nuclear network, and time-dependent thermalization of decay products from individual $\alpha$ and $\beta^-$ decay reactions.
In contrast to expansion opacities and other wavelength-binned treatments, a line-by-line treatment enables us to include fluorescence effects and associate spectral features with the emitting and absorbing lines of individual elements.
We find variations in the synthetic observables with both the polar and azimuthal viewing angles.
The spectra exhibit blended features with strong interactions by Ce~III, Sr~II, Y~II, and Zr~II that vary with time and viewing direction. We demonstrate the importance of wavelength-calibration of atomic data using a model with calibrated Sr, Y, and Zr data, and find major differences in the resulting spectra, including a better agreement with AT2017gfo.
The synthetic spectra for near-polar inclination show a feature at around 8000~\AA, similar to AT2017gfo.
However, they evolve on a more rapid timescale, likely due to the low ejecta mass (0.005~$\Msun$) as we take into account only the early ejecta.
The comparatively featureless spectra for equatorial observers gives a tentative prediction that future observations of edge-on kilonovae will appear substantially different from AT2017gfo.
We also show that 1D models obtained by spherically averaging the 3D ejecta lead to dramatically different direction-integrated luminosities and spectra compared to full 3D calculations.
\end{abstract}

\keywords{Neutron stars --- nuclear astrophysics --- r-process --- transient sources --- gravitational wave astronomy --- radiative transfer simulations}

\section{Introduction}
The gravitational wave event GW170817 from a neutron star merger (NSM) was complemented by a rich set of electromagnetic signals \citep{Abbott:2017it} including the associated kilonova AT2017gfo \citep{Smartt:2017kw,Villar:2017ct}.
Simple analysis of the luminosity decline rate of AT2017gfo shows consistency with radioactively decaying material that has undergone the rapid neutron capture process (r-process), responsible for synthesizing many of the elements heavier than Fe \citep{Metzger:2010hp}.
Another indication that very heavy elements have been produced is the inferred high opacity of the ejecta \citep{Kasen:2017kk}, which is considered to be a signature for the presence of lanthanides or actinides \citep{Kasen:2013dh}.

The observed time series of spectra and the luminosity evolution of AT2017gfo have provided a powerful set of constraints for testing theoretical NSM models, the high-density Equation of State (EoS), and r-process nucleosynthesis.
However, linking the electromagnetic observations back to the underlying physical conditions requires accurate models of radiative transfer in the ejecta, which depend on accurate atomic data.

The availability of atomic data from experiments is too limited to constitute a dataset for radiative transfer calculations.
Instead, systematic atomic structure calculations for trans-Fe elements (including lanthanides) have been performed by \citet{Fontes:2020jg} and \citet{Tanaka:2020gp}, and both provide tables of wavelength-binned opacities (i.e., opacities obtained as averages over discrete wavelength intervals).
In addition, the energy levels and transition data of \citet{Tanaka:2020gp} are available via the Japan-Lithuania Opacity Database for Kilonova\footnote{\href{http://dpc.nifs.ac.jp/DB/Opacity-Database/}{http://dpc.nifs.ac.jp/DB/Opacity-Database/}} \citep{Kato:2021ud}, which enables the calculation of individual line opacities.
However, the energy levels of \citet{Tanaka:2020gp} have not been calibrated to be consistent with observed transition lines \citep[as noted by][]{Domoto:2021gh}, which can severely affect the locations of spectral features and hinder the comparison between synthetic and observed spectra.
For some elements and ionization stages, calibrated atomic data is available \citep[e.g., ][]{Kurucz:2018vd}, but the coverage of the heavy elements is incomplete.

To date, most radiative transfer simulations for kilonovae use simplified ejecta models, such as with analytical density structures \citep{Metzger:2010hp,Kasen:2013dh,Bulla:2019ik,Even:2020cm,Banerjee:2020gp,Korobkin:2021fl,Domoto:2021gh,Wollaeger:2021iv,Gillanders:2022dz,Pognan:2022dn,Pognan:2022bp}.
With recent claims of observational constraints on the ejecta geometry \citep[e.g.,][]{Sneppen:2023gi}, it is especially important to understand whether multi-dimensional simulations can be reconciled with observations.
While there are studies that consider NSM ejecta distributions directly from simulations in two dimensions \citep{Kasen:2015hc,Kawaguchi:2018db,Korobkin:2021fl,Kawaguchi:2021hw,Bulla:2022ws,Just:2022ev,Kawaguchi:2022bt} or three dimensions \citep{Tanaka:2013dq,Collins:2023ha,Neuweiler:2023dd}, these calculations use some form of simplified opacity treatment, generally by combining line transitions into wavelength bins rather than calculating individual line-by-line opacities.

While line-by-line opacities are widely used for supernova calculations, the same method applied to kilonovae becomes computationally more expensive due to the substantial abundances of trans-Fe heavy elements, which have complex electronic structures that lead to vast numbers of energy levels and transition lines \citep{Kasen:2013dh}. Published kilonova calculations with line-by-line opacities are limited to simplified ejecta models and often a small selection of elements.
For example, \citet{Pognan:2022dn,Pognan:2022bp} apply line-by-line opacities in their non-LTE investigation of Te, Ce, Pt, and Th with a spherically-symmetric ejecta model but do not predict synthetic spectra. \citet{Gillanders:2022dz} present synthetic spectra calculated with line-by-line opacities for an extensive set of atomic data from H to U, but assume spherically symmetric ejecta with an inner photospheric boundary.

Using wavelength-binned opacities instead of line-by-line opacities has consequences beyond merely decreasing the wavelength resolution of the resulting synthetic spectra.
With binned opacities, the absorbing atomic transitions are unknown in the radiative transfer calculation, and re-radiation following absorption is typically handled either by assuming perfect scattering (no wavelength change) or perfect thermalization (emission according to the Planck function).
In order to avoid this simplification and apply a detailed treatment of the subsequent de-excitation radiation \citep[including fluorescence, ][]{Lucy:2002hw}, it is necessary to associate each bound-bound absorption with a specific transition.
When emission and absorption are associated with specific transitions, features in the synthetic spectra can be directly linked with particular elements, which would otherwise (i.e., with a wavelength-binned scheme) have to be inferred based on estimated line strengths or a comparison of several calculations with different compositions \citep[e.g.,][]{Domoto:2021gh,Sneppen:2023ks}.

The only direct comparison between synthetic spectra calculated using line-by-line opacities, line-binned opacities, and expansion opacities is made by \citet{Fontes:2020jg} for a one-dimensional test problem with a pure-Nd composition, and neglecting fluorescence effects.
Even this simple model leads to differences in resulting spectra between line-by-line and binned opacities \citep[see][figure 13]{Fontes:2020jg}.

In this work, we present 3D radiative-transfer calculations that include, for the first time, a line-by-line Sobolev treatment for tens of millions of lines based on three-dimensional ejecta self-consistently obtained from NSM simulations with a nuclear network. We show that our forward modeling approach can produce a sequence of spectra in the polar direction that look similar to AT2017gfo, and predict relatively featureless spectra for equatorial observers. We also associate spectral features with particular elements in the ejecta composition and demonstrate the importance of using multidimensional models.

\section{Method}

\subsection{NSM ejecta model and r-process calculation}\label{sec:nsm}
Our simulation of two merging 1.35 M$_\odot$ neutron stars adopts the SFHo equation of state \citep{Steiner:2013hi} and has been calculated with a 3D general relativistic smoothed-particle hydrodynamics (SPH) code \citep{Oechslin:2002iu,Bauswein:2010fu} employing the conformal flatness condition \citep{Isenberg:1980wj,Wilson:1996gr} and including a recently-developed leakage-plus-absorption neutrino treatment \citep[ILEAS,][]{ArdevolPulpillo:2019ea}. 

Since the SPH simulation terminates at 20 ms after the merger, we only capture the early, dynamical ejecta, which in our case has a mass of 0.005 \Msun and is made up of $\sim$2000 SPH particles.
The same NSM simulation was used in three-dimensional radiative transfer calculations with wavelength-independent (gray) opacities by \citet{Collins:2023ha}.

For performing 3D frequency-dependent radiative transfer calculations with \artis\footnote{\href{https://github.com/artis-mcrt/artis/}{https://github.com/artis-mcrt/artis/}\\The C++ source code is provided as is with no support or documentation under the BSD 3-Clause License.} \citep{Sim:2007is,Kromer:2009hv,Shingles:2020gy}, we require a snapshot of the density, composition, and thermal energy of each cell in a Cartesian velocity grid.
We use the same SPH mapping technique as \citet{Collins:2023ha}, in which particles are propagated with constant velocity for 0.5 s after the SPH simulation terminates, except that our grid size is $50^3$ (instead of $128^3$).
The SPH densities are then mapped onto the Cartesian velocity grid spanning -0.48c to 0.48c along each axis, from which homologous expansion is assumed (i.e., $\rho \propto t^{-3}$, see \citealt{Neuweiler:2023dd} for a study on the homologous assumption for dynamical NSM ejecta).

The first few minutes after a NSM involve a complex set of r-process reactions, so we use a nuclear network calculation \citep[as employed previously by][]{Collins:2023ha} to model the early heating and composition evolution.
The network calculation is run separately along the trajectory of each SPH particle, starting once the temperature falls below 10~GK and continuing beyond the end of the SPH simulation by assuming homologous expansion.

Synthetic light curves and spectra may depend sensitively on the assumptions and approximations used for the heating and thermalization treatment \citep[e.g.,][]{Barnes:2016ww,Bulla:2022ws}.
Although the network calculation gives summarized energy production rates that can be used for approximate heating \citep[e.g.,][]{Collins:2023ha}, a detailed treatment of thermalization requires complete information about the energies of emitted $\gamma$-rays and charged particles, so we extend \artis to treat individual $\alpha$ and $\beta^-$ decay reactions.
We take a snapshot of the network calculations at 0.1~days to ensure that capture reactions have slowed sufficiently such that only $\alpha$ and $\beta^-$ decay reactions are significant (see \autoref{appendix:decayvalidation} for comparison with the full network calculations).
We then map the partial density of each nuclide from the particles to the Cartesian grid in a similar way as the total mass density. The network calculations also give the integrated energy released in each cell (excluding neutron losses and assuming adiabatic expansion) from all reactions prior to the snapshot time, which \artis uses to seed thermal energy packets and set the initial cell temperatures.

For comparison with the full 3D model, we also construct a 1D ejecta model by spherically-averaging and binning the 3D ejecta into 25 radial velocity shells (uniformly spaced from $v_{rad}=0$ to $v_{rad}=0.48c$) with densities chosen such that the mass, as well as the partial masses of individual nuclides, are preserved within each shell.

\subsection{Thermalization of decay products}

The tracking of individual decay reactions in \artis enables the use of per-decay tabulated $\gamma$-ray spectra (in contrast to the single averaged $\gamma$ emission spectrum used by \citealt{Collins:2023ha}), as well as a new time-dependent treatment of $\alpha$ and $\beta^-$ particle thermalization that uses emitted particle energies adapted to the local conditions in the ejecta.
The models presented here include 2,591 nuclides with $\alpha$ and $\beta^{-}$ reaction data ($\gamma$-ray spectra and mean emitted particle energies) from ENDV/B-VII.1 \citep{Chadwick:2011dq}\footnote{As provided by \href{https://github.com/hotokezaka/HeatingRate}{https://github.com/hotokezaka/HeatingRate}}.

For $\gamma$-rays, we use the same frequency-dependent transport scheme previously applied to Type Ia calculations \citep{Lucy:2005cx,Sim:2007is}. For $\alpha$ and $\beta^-$ particles, we assume that energy deposition takes place at the location of the particle emission, which is a plausible assumption given that magnetic fields may prevent particles from crossing a significant fraction of the ejecta before losing their energy \citep{Barnes:2016ww}.

Radioactively emitted particles lose energy over time to the ejecta until they reach equilibrium with the thermal pool. We adopt the approximation by \citet{Barnes:2016ww} for the energy loss rates of $\alpha$- and $\beta$-particles of $\dot{E}_\alpha=5\times10^{11}\rho$ and $\dot{E}_\beta=4\times10^{10}\rho$ [MeV/s], where $\rho$ is the density in units of g/cm$^3$.
Although this loss rate function is independent of energy, the loss rate is defined per particle, and therefore the total energy deposition rate depends on the number of particles yet to thermalize, which requires knowledge of individual radioactive decay rates and their associated emitted particle energies.
Even for the same total $\beta^-$ particle energy emission rate, energy from a large number of reactions that emit low-energy particles will thermalize on a shorter timescale than a slower rate of reactions that emit high-energy particles.
This treatment captures the local dependence of thermalization on both the 3D density structure and the composition (which determines the decay rates and energy spectrum of emitted particles) that are neglected in analytical formulae provided by \citet{Barnes:2016ww}, which are based on a uniform density model and used by the majority of existing kilonova models.
In \autoref{appendix:thermalization}, we show a comparison between our calculated thermalization efficiencies and the \citet{Barnes:2016ww} analytical formulae.

\subsection{Radiative transfer and atomic data}\label{sec:rtatomic}
We consider two different sets of atomic data, AD1 and AD2. The full atomic dataset for the AD1 models includes 71 elements, 283 ions, 194,035 energy levels and a total of 43.6 million bound-bound transitions. Levels and transitions for elements (C, O, Ne, Mg, Si, S, Ar, Ca, Fe, Co, Ni) are sourced from the CMFGEN compilation\footnote{Available at \url{http://kookaburra.phyast.pitt.edu/hillier/web/CMFGEN.htm}} \citep{Hillier:1990ux,Hillier:1998gw}, similar to the dataset used for the nebular Type~Ia models by \citet{Shingles:2020gy,Shingles:2022df} except that photoionization is not included. For elements Cu to Ra, we incorporate all ions and levels of the Japan-Lithuania database \citep{Tanaka:2020gp}, which covers ionization stages from neutral to triply-ionized. We do not include any atomic data for actinides, which we will investigate in a future study.

Since the level energies in the Japan-Lithuania database are not calibrated to known transition lines, the low-lying permitted Sr~II 5p $\rightarrow$ 4d transitions (likely identified in AT2017gfo by \citealt{Watson:2019ki}, but see \citealt{Tarumi:2023cd} for an alternative explanation for the same feature with He) occur at rest wavelengths of 17903, 18190, and 20291 \AA\ instead of 10036, 10327, and 10914 \AA\xspace \citep{Sansonetti:2012gr}.
To address this, we have separately produced calculations with a different atomic dataset called AD2 that is the same as AD1, except that the data for Sr, Y, and Zr are from the \citet{Kurucz:2018vd} extended line list, which includes calibrated wavelengths where available.

We simulate a time range of 0.1 to 80~days for 3D~AD2, although the validity of our synthetic observables (light curves and spectra) is limited by light travel time effects to between 0.18 and 13.6~days (see \autoref{appendix:simtime} for details).

The radiative transfer calculations assume local thermodynamic equilibrium (LTE) with Boltzmann level populations and ionization balance according to the Saha equation.
The LTE assumption should be reasonably accurate within the first few days (\citealt{Pognan:2022dn}, but see \citealt{Tarumi:2023cd} for possible NLTE effects on Sr and He) and beyond this time, we only consider the luminosity to be reliable, since it simply converges to the deposition rate in optically thin ejecta and no longer has a dependence on the excitation or ionization state.

\section{Results}
\subsection{Luminosity evolution}

\begin{figure}
    \begin{center}
    \includegraphics[width=0.48\textwidth]{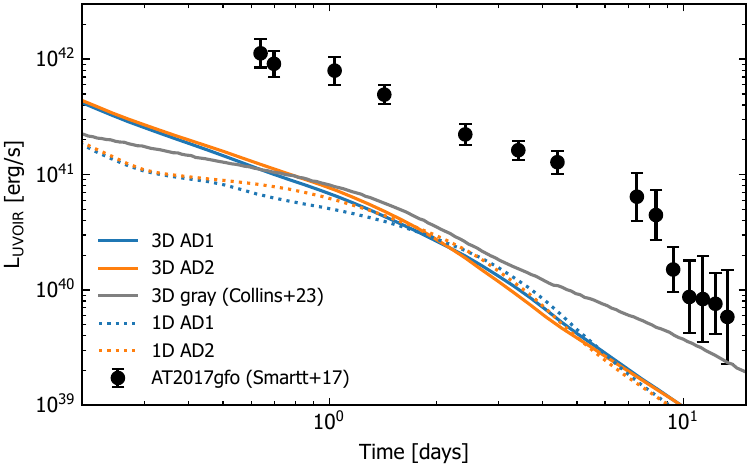}
    \end{center}
    \caption{Direction-integrated luminosity versus time for the models 3D~AD1, 3D~AD2, 1D~AD1, 1D~AD2, the 3D gray opacity model of \citet{Collins:2023ha}, and inferred bolometric luminosity of AT2017gfo \citep{Smartt:2017kw}.}
    \label{fig:lightcurves}
\end{figure}

\autoref{fig:lightcurves} shows the spherically-averaged light curves for all four models with 3D or 1D ejecta and AD1 or AD2 atomic datasets, the \citet{Collins:2023ha} 3D gray opacity calculations for the same NSM ejecta, as well as the inferred bolometric (all wavelengths) flux for AT2017gfo \citep{Smartt:2017kw}.

Compared to AT2017gfo, the bolometric luminosities of all of the models are lower by about a factor of ten, mainly because our model mass includes only early ejecta with a mass of 0.005 M$_\odot$, which is about ten times lower than the mass estimated for AT2017gfo \citep{Smartt:2017kw}.

Both the AD1 and AD2 atomic datasets lead to approximately equal total bolometric luminosities, which indicates that wavelength calibration of Sr, Y, and Zr does not significantly affect the overall optical depth of the ejecta.

The luminosities of the 3D models are higher than for the 1D models until about two days.
This is a consequence of the strong anisotropies in our NSM model.
A complex 3D ejecta structure with local opacity variations exhibits pathways of lower optical depth and therefore allows light to effectively escape faster than in a 1D configuration.
After a couple of days, when the expanding ejecta have become optically thin, the luminosities of the 1D and 3D models converge to the rate of thermalized beta particle energy.

The luminosities of the 3D~AD1 and 3D~AD2 models agree reasonably well with the 3D gray opacity calculation of \citet{Collins:2023ha} for the first couple of days.
This shows that the $Y_e$-dependent gray opacity tables of \citet{Tanaka:2020gp} are a good approximation to the opacities calculated line-by-line with the time- and location-dependent temperatures and composition of our model.
At times later than about two days, the thermalization efficiency for $\beta$ particles begins to decrease significantly and the luminosity of our model decreases more rapidly than the \citet{Collins:2023ha} model, which assumes fully-efficient $\beta^-$ thermalization.

Overall, we find that the use of 3D ejecta is critical for modeling the early (\textless 2~days, while optically thick) luminosity evolution, and the opacity treatment is comparatively less important.
This is also approximately the time range within which the assumption of fully-efficient thermalization of $\beta$ particles is valid.

\subsection{Observing direction maps}
\begin{figure}
    \begin{center}
    \includegraphics[width=0.48\textwidth]{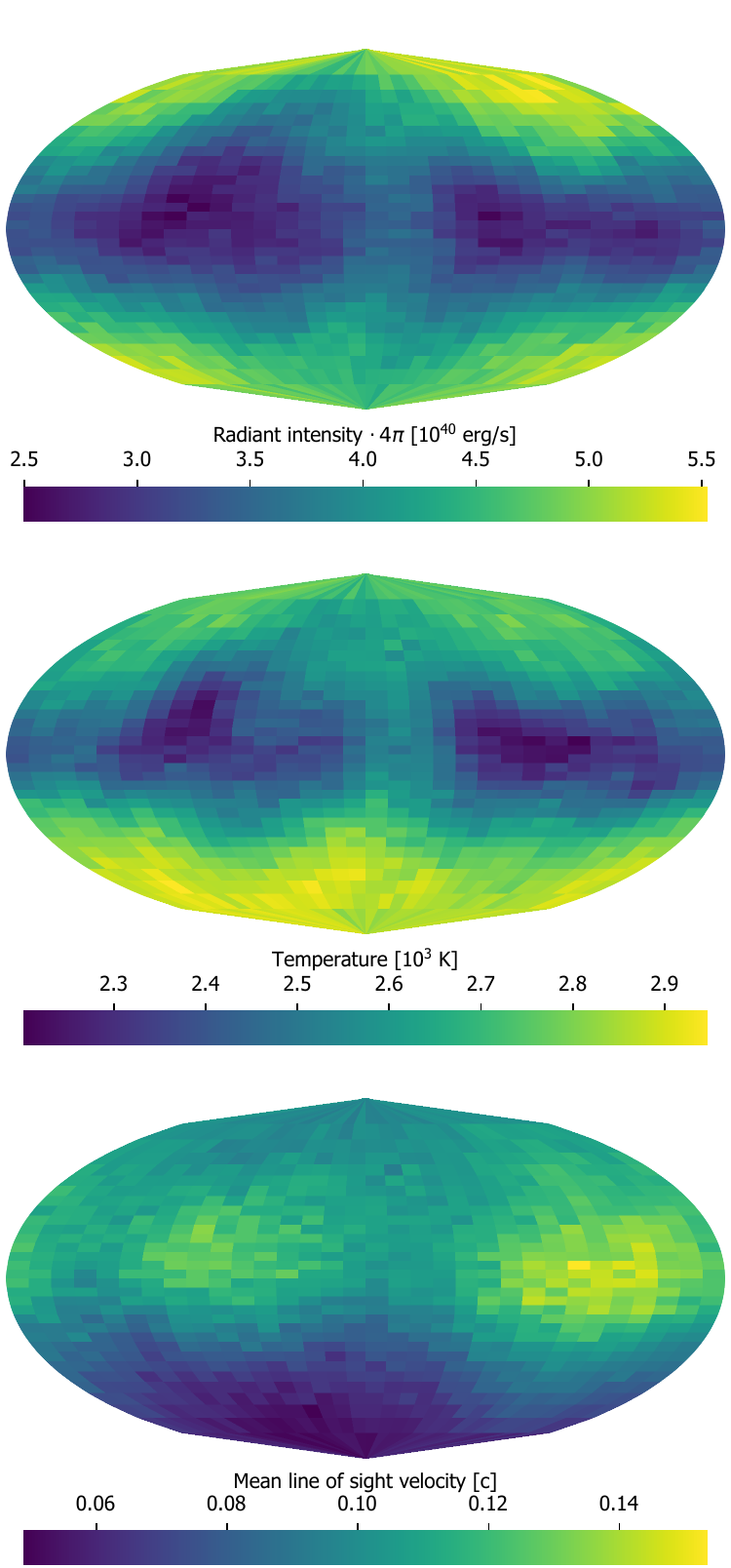}
    \end{center}
    \caption{Mollweide projections of direction-dependent quantities for 3D~AD2 UVOIR packets arriving at the observer between 1.5 and 1.8d: radiant intensity times 4$\pi$ solid angle, mean temperature at last interaction, and line of sight velocity at last interaction. For these figures, we use 32x32 direction bins, uniformly spaced in azimuthal angle (horizontal) and cosine of the polar angle (vertical) to give the same solid angle in each bin.}
    \label{fig:spherical}
\end{figure}

With 3D models, we are able to explore the direction dependence of the synthetic observables. \autoref{fig:spherical} shows Mollweide projections for the isotropic-equivalent luminosity (radiant intensity times $4\pi$), the mean temperature of the emission locations, and the mean line-of-sight velocity of emitting locations for light arriving at the observer between 1.5 and 1.8~days.

The faintest lines of sight are those viewing toward the equator (merger plane), where the model has the lowest $Y_e$ and a more lanthanide-rich composition.
The polar-angle variation of the composition \citep[e.g.,][]{Kullmann:2022cx,Collins:2023ha} leads to a higher mean last-interaction ejecta velocity for equatorial observers, where the ejecta density and temperature are lower than lower-velocity central regions.
Compared to polar observers, equatorial observers receive emission originating from regions with a higher average line-of-sight velocity (0.16--0.17c versus 0.10--0.13c), and lower temperature (2500--2800 K versus 3200-3400 K).

Apart from the variation with polar angle, which was already found in previous studies \citep[e.g.,][]{Just:2022ev,Neuweiler:2023dd}, we also see substantial variation in the emission as a function of the azimuthal angle.
Since most existing multi-dimensional kilonova studies assume 2D axisymmetry, they cannot capture the dependence on azimuthal angle. Our results suggest that this dependence can be important as well, at least for merger models similar to the one considered here.

\subsection{Spectra toward pole and equator}

\begin{figure*}[htb]
    \begin{center}
    \includegraphics[width=0.49\textwidth]{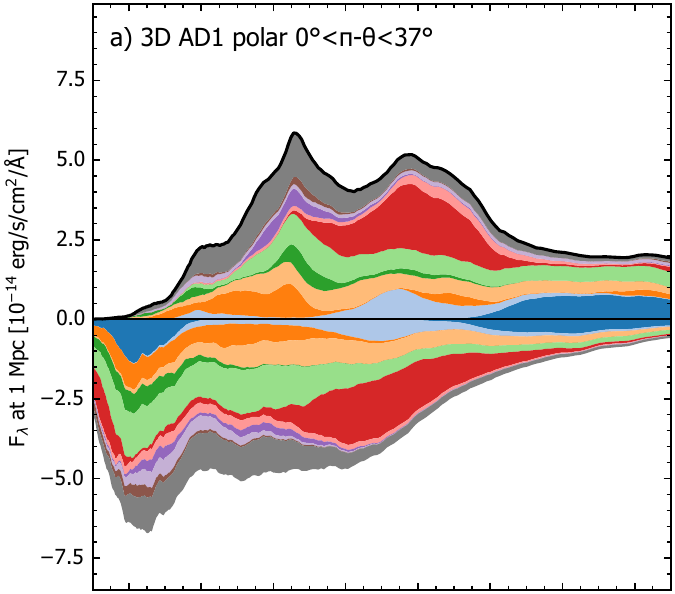}
    \includegraphics[width=0.49\textwidth]{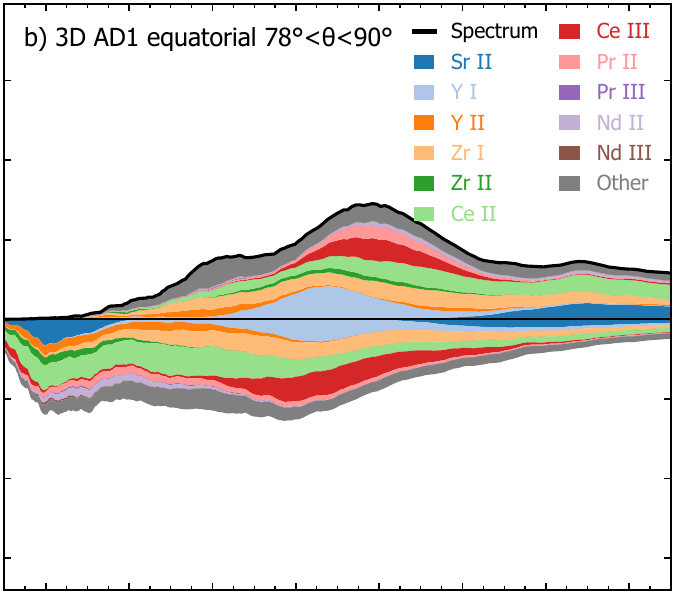}
    
    \includegraphics[width=0.49\textwidth]{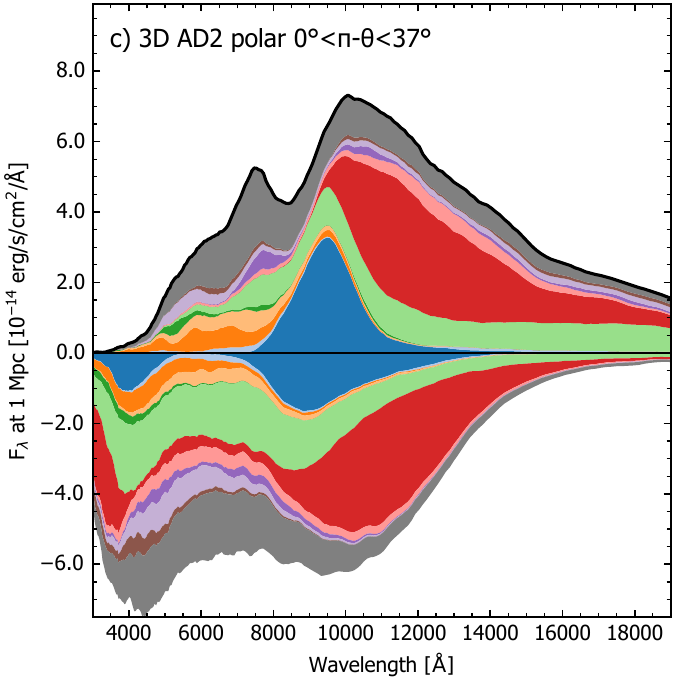}
    \includegraphics[width=0.49\textwidth]{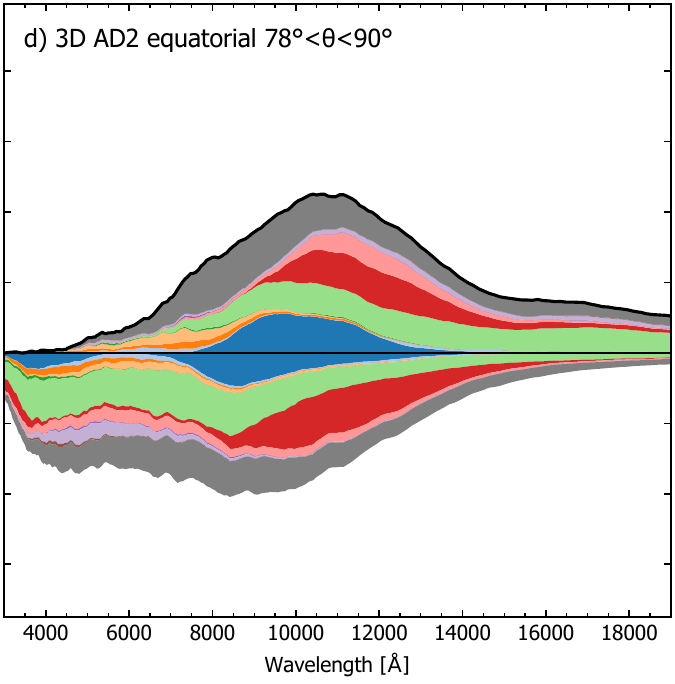}
    \end{center}
    
    \caption{Spectra for polar and equatorial viewing directions for the 3D~AD1 and 3D~AD2 models at 1.1~days. The height of each wavelength point is colored according to the emitting species of the last interactions of the emerging radiation packets. The area under the horizontal axis shows the distribution of frequencies (colored by absorbing/scattering ion) just prior to the last interactions of the emerging packets. The 11 most-significant ions are separately colored, while the "Other" group combines many smaller contributions from other ions.}
    \label{fig:spectra_directions}
\end{figure*}

\autoref{fig:spectra_directions} shows spectra at 1.1~days for the 3D~AD1 and 3D~AD2 models for polar and equatorial inclinations (averaged over azimuthal angle to reduce Monte Carlo noise).
At each wavelength, we divide the flux into the set of emitting ions based on the last interactions of the emerging radiation packets.
Below the horizontal axis, we show the distribution of absorption frequencies (in the observer frame) at the last interaction and associate this flux with the set of absorbing/scattering ions.
This visualization gives some indication of the ions that are important for shaping the spectrum, although it does not consider the full history of interactions experienced by the emerging packets.

We notice substantial variation in the spectra with both the polar viewing angle as well as the adopted atomic dataset.
For both 3D models, the equatorial spectra are fainter and show comparatively few pronounced features compared to the spectra for the near-polar inclination.
In contrast to supernova spectra, for which individual spectral features can often be attributed to the transitions of one or two ions, our model results suggest that the spectral formation in kilonovae is much more complex.
Across the wavelength range, the emerging packets of radiation have interacted with multiple ions.
Fluorescence by several elements (especially Sr, Y, Zr, and Ce) is highly effective at redistributing flux bluewards of $\sim$4000~\AA\ to longer (infrared) wavelengths.

The effect of wavelength calibration is dramatic for the Sr~II contributions (see \autoref{sec:rtatomic} for details of the rest wavelength differences): In the 3D~AD1 spectra, an extremely broad Sr~II triplet feature exists around 15000-20000~\AA, while in the 3D~AD2 model, the same feature occurs at around 7500-12500~\AA, blue-shifted from the rest wavelengths of around 10000~\AA.
In the 3D~AD2 case, the wavelength-calibrated Sr~II transitions occur much closer to the peak of the continuum, and therefore contribute more strongly to the overall spectral distribution.
Absorption by Sr~II (below the axis) is further blue-shifted than the Sr~II emission, as would be expected from a P Cygni feature \citep{Watson:2019ki}.
However, we see that Sr~II also emits by fluorescence following the absorption of shorter-wavelength ($\sim$4000~\AA) radiation.

The complex interplay of ion interactions is also indicated by the degree to which the calibration of Sr, Y, and Zr has consequences for the interactions of other ions (e.g., Pr~III) that have identical atomic data in the AD1 and AD2 models.

We also see that Ce~III is often the last interaction of packets emerging around 10000-12000~\AA, while \citet{Domoto:2022wg} have shown that calibration would substantially weaken Ce~III lines in this range.
However, despite frequent interactions with Ce~III, a comparison with a separate simulation performed without Ce (but otherwise identical to 3D~AD2) suggests that the Ce only has a modest impact on the shape of the spectral distribution, and interacts mainly by scattering, with some fluorescence of optical flux to infrared wavelengths.

\subsection{Spectral evolution}

\begin{figure*}[htb]
    \begin{center}
    \includegraphics[width=0.48\textwidth]{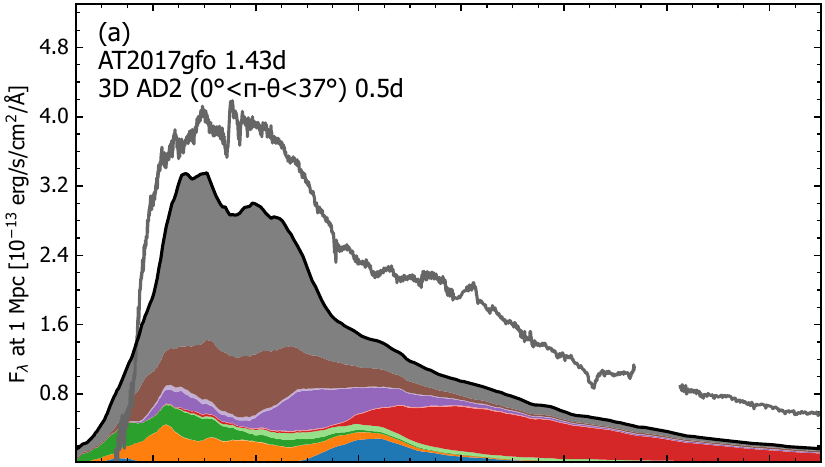}
    \includegraphics[width=0.48\textwidth]{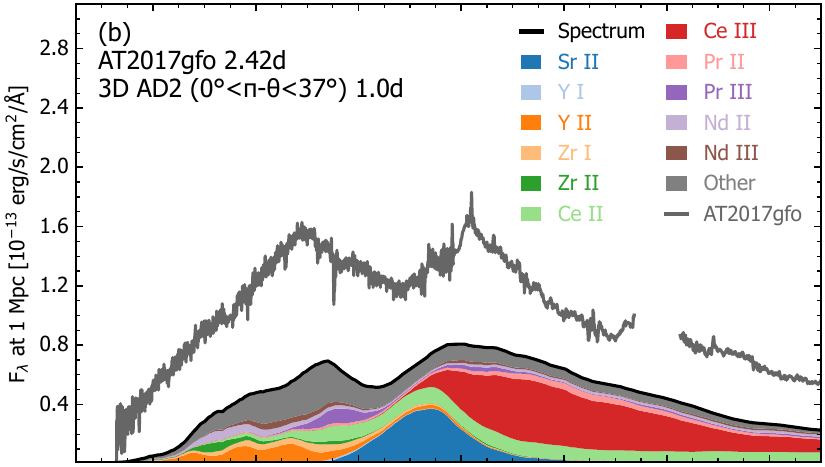}

    \includegraphics[width=0.48\textwidth]{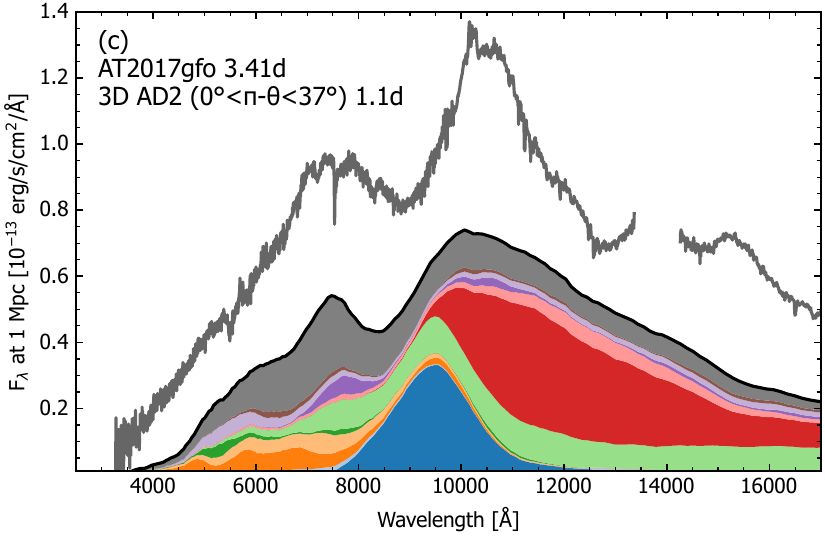}
    \includegraphics[width=0.48\textwidth]{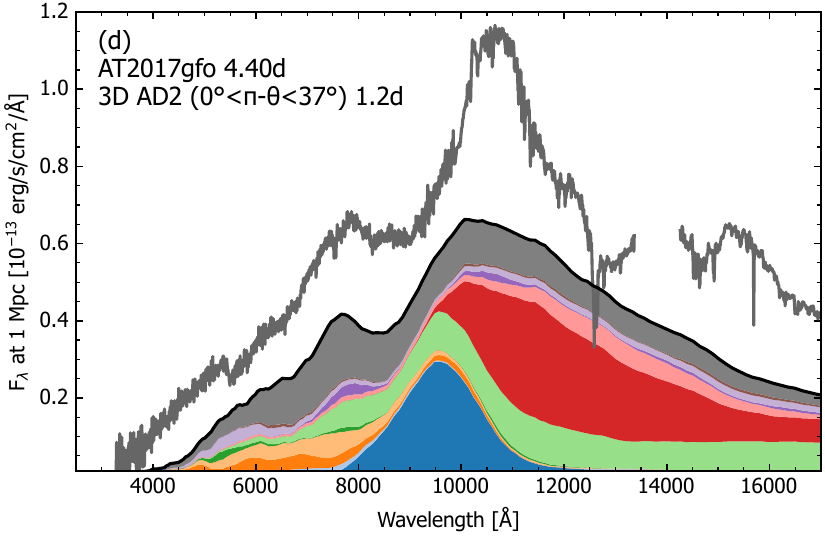}
    \end{center}
    \caption{Time series of spectra in the polar direction of the 3D~AD2 model compared to reddening and redshift corrected spectra of AT2017gfo \citep{Pian:2017ke,Smartt:2017kw}. The area under the spectra have been coloured by the emitting species of the last interactions of the emerging packets. The times of the \artis and AT2017gfo spectra intentionally do not match.}
    \label{fig:specsequence}
\end{figure*}

\autoref{fig:specsequence} shows a time series of spectra for the polar direction of the 3D~AD2 model and a series of AT2017gfo observations\footnote{Engrave data release available at \href{http://www.engrave-eso.org/AT2017gfo-Data-Release/}{http://www.engrave-eso.org/AT2017gfo-Data-Release/}} \citep{Pian:2017ke,Smartt:2017kw}.
We select a viewing angle bin closest to the +z pole ($\pi$-$\theta \leq 37$\textdegree, where $\theta$ is the angle from the positive z-axis, as defined by \citealt{Collins:2023ha}) containing the estimated viewing angle to AT2017gfo of between 14\textdegree\ and 28\textdegree\ \citep{Mooley:2022kw}.

Although the model evolves about 3-4 times quicker than AT2017gfo, the 3D~AD2 spectra pass through phases that are remarkably similar to the observed spectra.
The evolution being too rapid could be a consequence of the low-mass of our model that includes only the early ejecta, as a larger mass would increase the diffusion timescale.

The 3D~AD2 polar spectrum at 0.5~days has a single broad bump at around 4000-7000~\AA, which is similar to AT2017gfo at 1.4~days.
In our model, the broad peak is a blend, where the Monte Carlo packets have had notable interactions with Zr~II, Y~II, Nd~III, and Pr~III, as well as interactions with many other ions.

At 1.0~days, the 3D~AD2 spectrum begins to exhibit more pronounced spectral features, with similarities to the AT2017gfo spectrum at 2.42~days.
However, the spectral features in the model occur at shorter wavelengths (by up to $\sim$1000~\AA) which might suggest that the expansion velocities in the ejecta model are too high.

At 1.1~days, the distribution has shifted further to the red, with stronger features and a similar spectrum to AT2017gfo at 3.41~days.
There are spectral features in AT2017gfo redward of 12000~\AA\ that are not reproduced by our model (possibly due to our use of uncalibrated La~III and Ce~III lines, see \citealt{Domoto:2022wg} and \citealt{Tanaka:2023cf}), but the approximate agreement with the polar spectrum is interesting when contrasted with the predicted spectra in the equatorial direction.
The equatorial spectrum of 3D~AD2 at 1.1~days (shown in \autoref{fig:spectra_directions}) is comparatively lacking in pronounced spectral features and is much less similar to the AT2017gfo spectrum.
The closer spectral match in the polar direction of our model independently supports the polar inclination of AT2017gfo, which has been previously inferred with other methods that involve different assumptions about the merger \citep{Mooley:2022kw}.

At 1.2~days, the 3D~AD2 spectral energy distribution shifts further toward redder wavelengths, with the overall spectrum appearing similar to AT2017gfo at 4.40~days.

These results show that forward modeling of a merger simulation that has not been tuned to match AT2017gfo, nevertheless shows remarkable similarities with observations when viewed in the polar direction.

\subsection{Spherically-averaged ejecta models}\label{sec:1d3d}

\begin{figure}
    \begin{center}
    \includegraphics[width=0.49\textwidth]{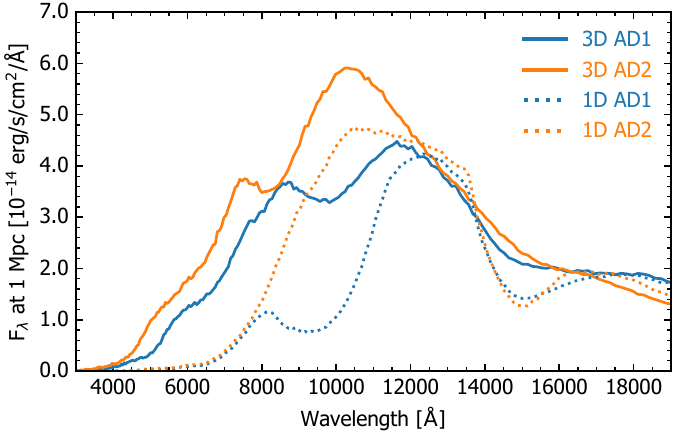}
    \end{center}
    \caption{Spherically-averaged spectra at 1.1~days for the 3D~AD1 (solid blue), 3D~AD2 (solid orange), 1D~AD1 (dashed blue), 1D~AD2 (dashed orange) models.}
    \label{fig:spectra_angleavg_all3d1d}
\end{figure}

\autoref{fig:spectra_angleavg_all3d1d} shows the spectra at 1.1~days averaged over all viewing directions for the models calculated in either full 3D, or with 1D spherically-averaged ejecta with either the AD1 or AD2 atomic datasets.
Even when the same atomic dataset is used, the 1D calculations predict very different spectral energy distributions compared to the 3D calculations.
In general, the 3D models produce substantially more flux shortward of $\sim$8000 \AA, which highlights the need for multidimensional models when predicting spectral energy distributions.

For the AD1 case, the 1D and 3D models produce very different sets of spectral features. Indeed, the 3D~AD1 model shows a flux maximum at $\sim$7000 \AA, while 1D~AD1 has a maximum at around 12000 \AA.
For the AD2 case, the 1D and 3D models each produce a spectral maximum at around 10000 \AA, although the 1D model also shows a second distinct feature at around 12000--13000 \AA\ that is not present for the 3D model.

\section{Discussion and Conclusions}
We have presented the so-far most self-consistent radiative-transfer calculations of NSM ejecta, taking into account the full 3D spatial and time dependence of the density, composition, heating rate, and thermalization efficiency, and using for the first time, a line-by-line opacity treatment.
The line-by-line treatment enables the handling of fluorescence and allows the association of spectral features with individual atomic transitions.
We have shown that the Sr~II triplet (5p $\rightarrow$ 4d) plays an important role in shaping the spectrum around 8000 to 11000~\AA, as has been suggested on the basis of simplified models \citep[e.g.,][]{Watson:2019ki}, which is an important confirmation given the substantial differences between 1D and 3D model results presented in \autoref{sec:1d3d}.
However, we show that important contributions to the spectra are also made by Ce~III \citep[although our Ce~III lines are uncalibrated, see][]{Domoto:2022wg}, Zr~I, Zr~II, and Y~II \citep[as suggested from observations by][]{Sneppen:2023ks}.

We have produced a time series of viewing-angle-dependent spectra that, in the polar direction, are similar to AT2017gfo, and which show distinctly more features compared to equatorial spectra.
These results independently support the inferred polar inclination of AT2017gfo and suggest the possibility that distinctly different spectra will be obtained from future kilonovae observed at different inclination angles.
While we can only broadly constrain the inclination with our current model, future theoretical models offer the possibility of further constraining the inclination, which is highly desirable for gravitational-wave data analysis, distance estimates, Hubble constant inference, and the interpretation of NSM observations in general.

We obtain dramatically different spectra between the 1D and the 3D ejecta models derived from the same NSM model, despite keeping all other aspects of the radiative transfer method identical.
In particular, there are strong observer-orientation effects (see \autoref{fig:spectra_directions}), and moreover, the spectra from a spherically-averaged 1D model do not resemble those of the underlying 3D model in any viewing angle.
We contend that this demonstrates the need for multidimensional models of kilonovae, even when the direction-dependence of the observables is not required.
This has important implications for interpreting analyses based on 1D empirical models \citep[e.g.,][]{Gillanders:2022dz}.

While the use of advanced treatments of radioactive decay and thermalization, atomic absorption and emission help to reduce systematic uncertainties in the radiative transfer calculations, the results of our study also highlight the importance of using accurate atomic data.
We have shown the importance of calibrating energy levels in atomic structure calculations to observed transition wavelengths, with major differences in the resulting synthetic spectra being produced when calibrated atomic data are used for Sr, Y, and Zr.
There are many heavy ions for which no calibrated atomic data are published, while work in this direction is ongoing \citep[e.g.,][]{Flors:2023dx}.
Future applications of calibrated data to radiative transfer calculations are likely to help in explaining additional features of kilonova spectra and correlating these with merger dynamics and remnant properties.

\begin{acknowledgments}
The authors thank the anonymous referee for providing helpful comments, as well as Brian Metzger for providing useful feedback.
AF, GL, GMP, LJS, and ZX acknowledge support by the European Research Council (ERC) under the European Union’s Horizon 2020 research and innovation program (ERC Advanced Grant KILONOVA No. 885281).
AB, OJ, CEC, and VV acknowledge support by the European Research Council (ERC) under the European Union’s Horizon 2020 research and innovation program under grant agreement No. 759253.
AB, CEC, AF, OJ, GL, GMP, LJS, and ZX acknowledge support by Deutsche Forschungsgemeinschaft (DFG, German Research Foundation) - Project-ID 279384907 - SFB 1245 and MA 4248/3-1.
AB and VV acknowledge support by DFG - Project-ID 138713538 - SFB 881 (“The Milky Way System”, subproject A10).
AB, CEC, AF, GL, GMP, OJ, LJS, and ZX acknowledge support by the State of Hesse within the Cluster Project ELEMENTS.
The work of SAS was supported by the Science and Technology Facilities Council [grant numbers ST/P000312/1, ST/T000198/1, ST/X00094X/1].
This work was performed using the Cambridge Service for Data Driven Discovery (CSD3), part of which is operated by the University of Cambridge Research Computing on behalf of the STFC DiRAC HPC Facility (www.dirac.ac.uk). The DiRAC component of CSD3 was funded by BEIS capital funding via STFC capital grants ST/P002307/1 and ST/R002452/1 and STFC operations grant ST/R00689X/1. DiRAC is part of the National e-Infrastructure.
The authors gratefully acknowledge the Gauss Centre for Supercomputing e.V. (www.gauss-centre.eu) for funding this project by providing computing time through the John von Neumann Institute for Computing (NIC) on the GCS Supercomputer JUWELS \citep{JulichSupercomputingCentre:2021ea} at Jülich Supercomputing Centre (JSC) and the VIRGO cluster at GSI for computational support.

\end{acknowledgments}

%

\vspace{5mm}


\software{
\textsc{NumPy} \citep{Oliphant:2007dm},
\textsc{Matplotlib} \citep{Hunter:2007ih},
\textsc{artis}\footnote{\href{https://github.com/artis-mcrt/artis}{https://github.com/artis-mcrt/artis}} \citep{Sim:2007is,Kromer:2009hv,Shingles:2020gy}, and
\textsc{artistools}\footnote{\href{https://github.com/artis-mcrt/artistools}{https://github.com/artis-mcrt/artistools}} \citep{Shingles_artistools_2023}}



\appendix

\section{ARTIS calculation and time range of synthetic observables}\label{appendix:simtime}
We use \artis version v2025.02.17 \citep{artis_2025}. The 3D~AD2 model (0.1 to 80~days) with $10^8$ Monte Carlo energy packets used 83~hours of wallclock time on 960 AMD EPYC 9654 Zen 4 CPU cores (79~kilo-core-hours).
The memory usage was less than 2~GB per core.

We simulate a time range of 0.1 to 80~days, although the range of validity of our synthetic observables is limited by light travel time to between 0.18 and 13.6~days.

Packets of radiation can escape from the simulation volume up to the end of the simulated time range, but the escape surface at the boundary of the simulation expands with constant velocity ($0.48c$ per axis, $0.83c$ at the corners), and packets might have already traveled some distance toward the observer before escaping.
The light-crossing time of the component of distance along the line of sight to the observer ($\vec{r} \cdot \hat{v}/c$) must be subtracted from the escape time to get a consistent time of arrival for the observer that is independent of the arbitrary escape surface.

The greatest time correction will occur for a packet moving in a radial direction and escaping at a corner of the simulation at the end time ($|\vec{r}_\mathrm{corner}| = 0.83c \times 80$~days).
In this case, the observer arrival time is less than the escape time by $(\vec{r} \cdot \hat{v}) / c = 0.83c \times 80$~days $= 66.4$~days.
The latest observer time for which we have complete coverage of escaping packets is then $80-66.4=13.6$~days.

For the earliest observation times, a packet emitted radially inwards from a simulation corner must have time to reach the origin (beyond which further travel time will be subtracted), which is $0.8c \times 0.1$~days, or 0.08~days after the 0.1~day simulation start time.
Therefore the synthetic observables are valid from 0.18 to 13.6~days (or to 3.4~days), while internal quantities such as deposition and temperatures are defined for the entire simulation time range.


\section{Validation of the ARTIS nuclear decay treatment}\label{appendix:decayvalidation}

The \artis treatment of $\alpha$ and $\beta^-$ decays uses a fast analytical calculation. With the assumption of pure decays (i.e., no capture reactions), there are no loops in the nuclear network, and therefore we can enumerate each of the finite number of paths that begin at a nuclide in the snapshot composition.
Given the snapshot abundances at time $t_s$, a modified \citet{Bateman:1910ue} equation (incorporating branching factors for each decay type) gives each path's contribution to the abundances of the nuclide at the end of the path at any time $t > t_s$. The abundance of any nuclide is then given by the sum of the contributions from all associated decay paths.

Here, we show that (with our chosen snapshot time of $t_s=0.1$~days) the decay treatment in \artis maintains close agreement with the iterative matrix-based calculation that includes capture reactions and spontaneous fission.

\begin{figure}
    \begin{center}
    \includegraphics[width=0.7\textwidth]{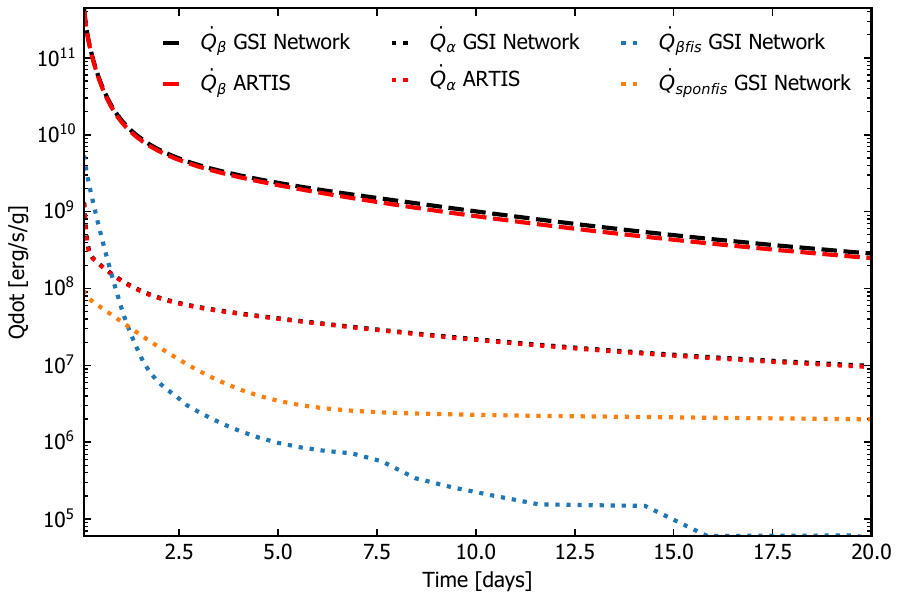}
    \end{center}
    \caption{Comparison of energy release rates of the \artis treatment in 3D~AD2 versus the output of the full nuclear network (GSI network). Neglecting fission reactions has only a negligible effect on the total energy release rates.}
    \label{fig:decaytracking}
\end{figure}

\autoref{fig:decaytracking} shows the global energy release rates as a function of time for both \artis and the full nuclear network calculation. The \artis treatment is able to match the evolution of $\dot{Q}_\alpha$ and $\dot{Q}_\beta$ closely, which are the total (including neutrino) energy release rates for the $\alpha$ and $\beta^-$ decays, respectively. The total $\dot{Q}$ is not shown because it is extremely close to the $\dot{Q}_\beta$ value.

\begin{figure}
    \begin{center}
    \includegraphics[width=0.48\textwidth]{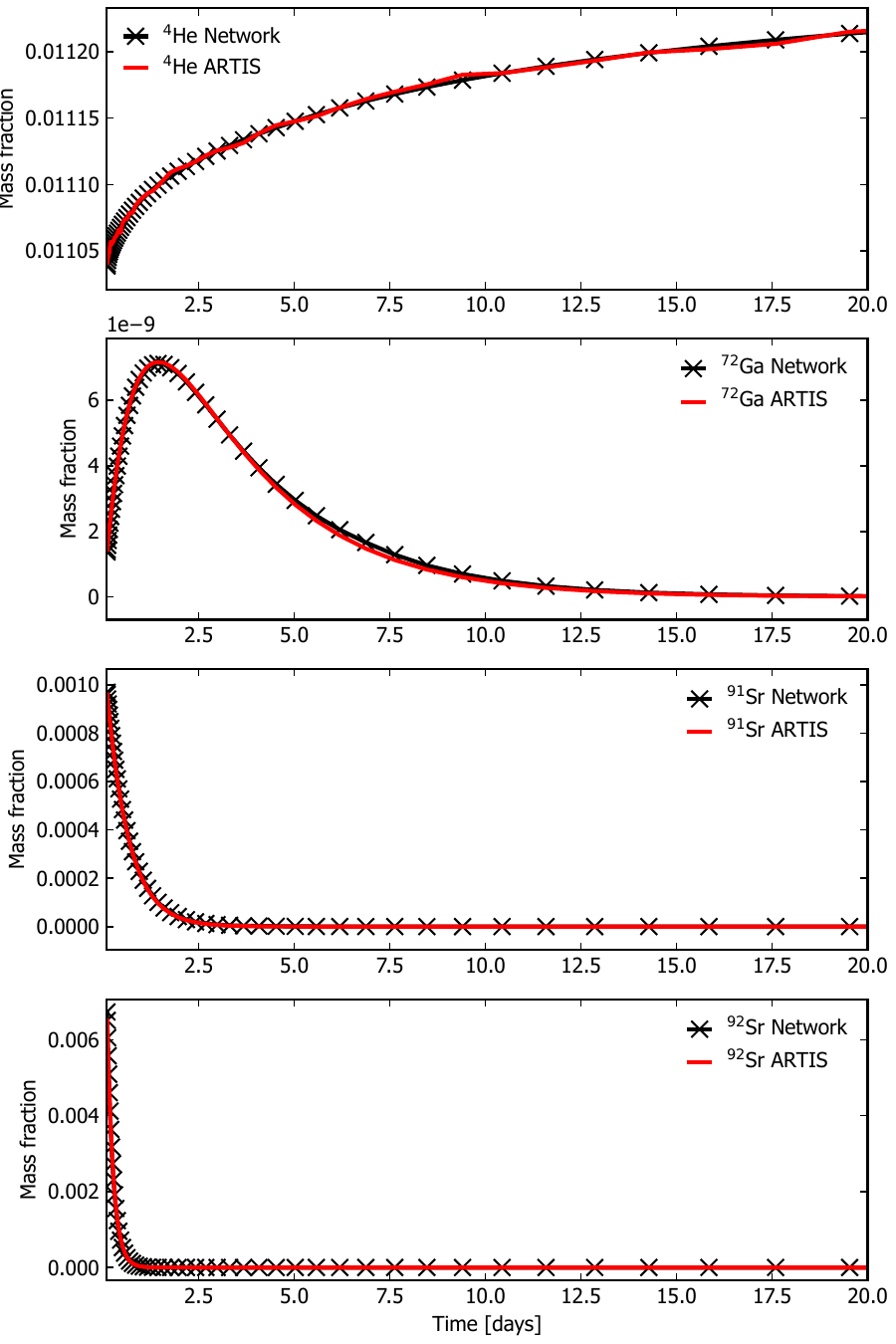}
    \includegraphics[width=0.48\textwidth]{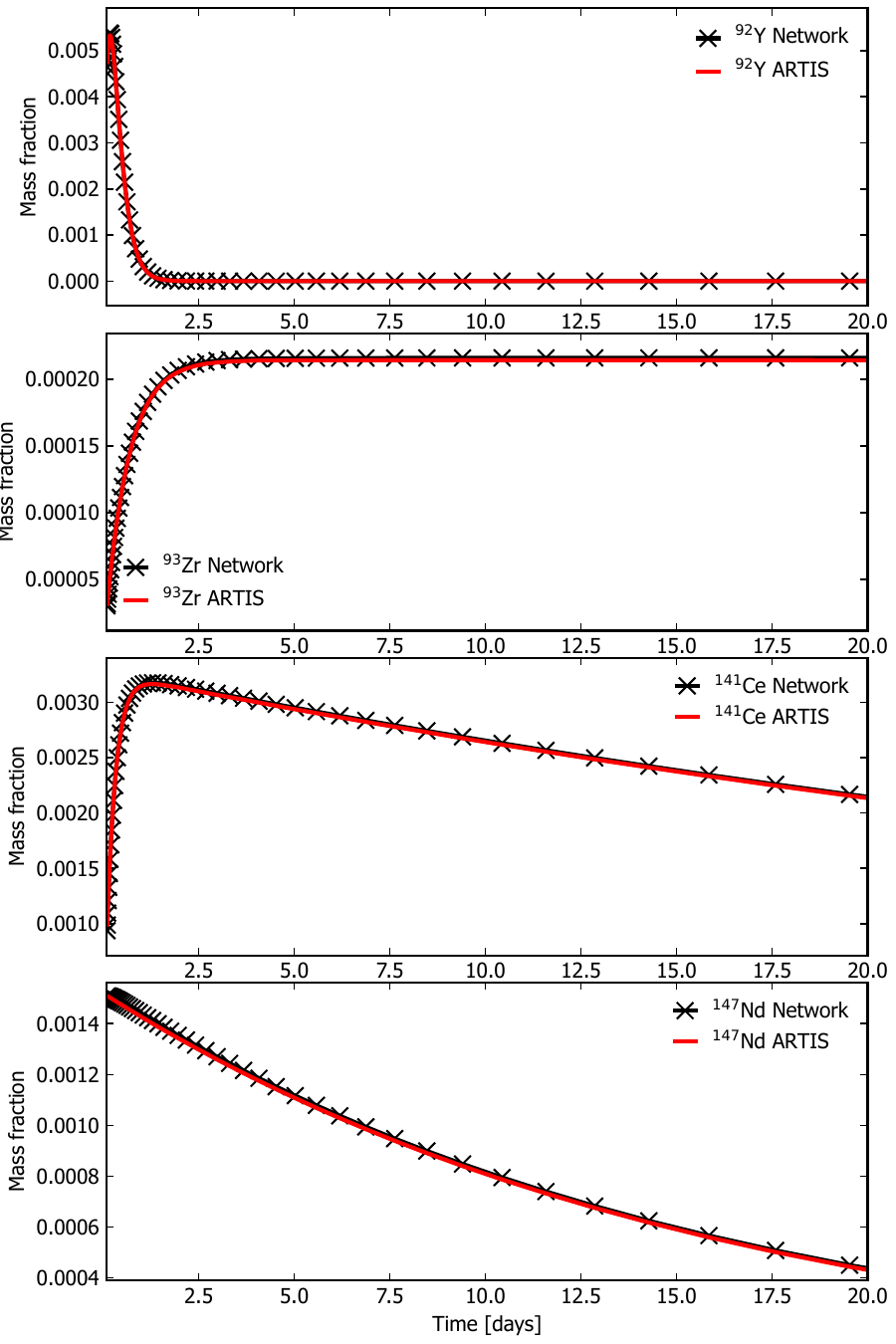}
    \end{center}
    \caption{Comparison between the nuclear mass fractions from the \artis decay-only treatment and the full nuclear network calculation for a cell at $z=0$ and radial velocity $v_r=0.24c$ with $Y_e=0.13$ .}
    \label{fig:abundtracking}
\end{figure}

\autoref{fig:abundtracking} shows a comparison of the abundance evolution for selected nuclides in both the \artis decay-only treatment and the full nuclear network calculation.
The abundances maintain close agreement throughout the time range, and in particular, the $^4$He abundance increases due to many $\alpha$ decays.

\section{Deposition and thermalization}\label{appendix:thermalization}
\begin{figure}
    \begin{center}
    \includegraphics[width=0.49\textwidth]{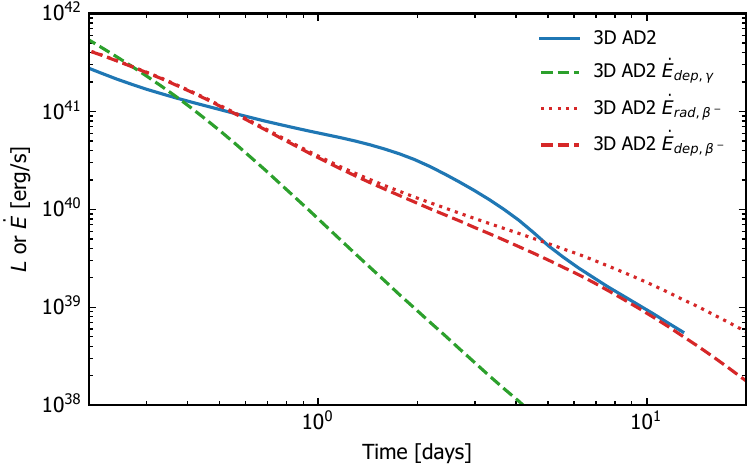}
    \includegraphics[width=0.49\textwidth]{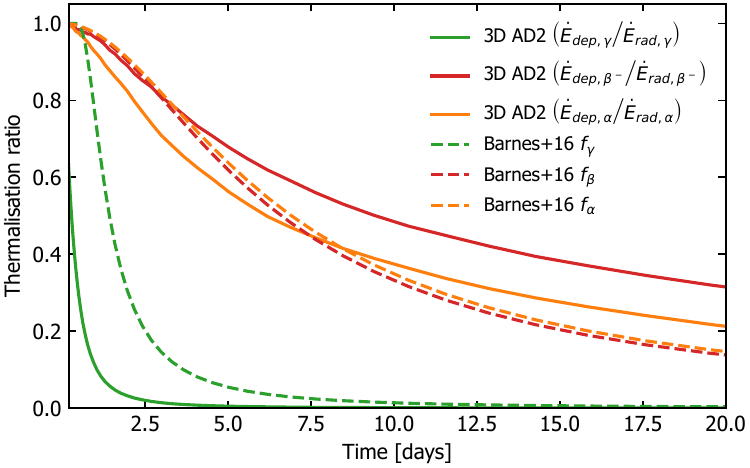}
    \end{center}
    \caption{Left panel: synthetic direction-integrated bolometric luminosity (all wavelengths excluding $\gamma$-rays), global deposition rate for $\gamma$-rays, and emission and deposition rates for $\beta^{-}$ particles for the 3D~AD2 model.
    Right panel: global thermalization ratios (power deposited divided by emitted) versus time for $\gamma$-rays, $\beta^{-}$, and $\alpha$ particles in the 3D~AD2 \artis model. We also show the \citet{Barnes:2016ww} analytical fits (Equations \ref{eq:fparticle} and \ref{eq:fgamma}) for comparison.}
    \label{fig:lightcurve_deposition}
\end{figure}

The left panel of \autoref{fig:lightcurve_deposition} shows the direction-integrated bolometric (ultraviolet, optical, and infrared) luminosity, as well as the global rates of energy deposited by $\gamma$-rays, and the energy released and deposited by $\beta$ particles for the 3D~AD2 model.
The right panel of \autoref{fig:lightcurve_deposition} shows the time evolution of the thermalization ratios for $\gamma$-rays, $\beta^{-}$, and $\alpha$ particles in \artis and for comparison, the \citet{Barnes:2016ww} analytical fits using the total mass and average velocity of our ejecta model.

The approximate thermalization efficiencies for $\gamma$-rays and massive particles are given by equations 32 and 33 of \citet{Barnes:2016ww},
\begin{align}
    f_p(t) &= \frac{\ln\left[1+2\left(\frac{t}{t_{\text{ineff,p}}}\right)^2\right]}{2\left(\frac{t}{t_{\text{ineff,p}}}\right)^2}\quad \text{and}\label{eq:fparticle}\\
    f_\gamma(t) &= 1 - \exp\left[ -\left(\frac{t}{t_{\text{ineff},\gamma}}\right)^{-2}\right],\label{eq:fgamma}
\end{align}
respectively, where $t_\text{ineff}$ refers to the time at which $\gamma$-rays or particles fail to thermalize efficiently (when the thermalization timescale exceeds the expansion time).

The thermalization timescales are defined by equations 16, 17, 20, 25 of \citet{Barnes:2016ww}:
\begin{align}
    t_{\text{ineff},\gamma} &= 1.4 M_5^{1/2} v_2^{-1} \text{ days}\\
    t_{\text{ineff},\beta} &= 7.4 \left( \frac{E_{\beta,0}}{0.5\text{ MeV}} \right)^{-1/2} M_5^{1/2}v_2^{-3/2} \text{ days}\\
    t_{\text{ineff},\alpha} &= 4.3 \times 1.8 \left( \frac{E_{\alpha,0}}{6\text{ MeV}} \right)^{-1/2} M_5^{1/2}v_2^{-3/2} \text{ days},
\end{align}
where $M_5$ is the ejecta mass in units of $10^{-5}$ \Msun, and $v_2$ is the average velocity computed according to $v_2=\sqrt{\int{ v^2 dm}/\int{dm}}/(0.2c)$.
For the analytic values presented here, we use the suggested initial particle energies of $E_{\beta,0} = 0.5$~MeV and $E_{\alpha,0}=6$~MeV and, although the Barnes formulae are not intended to be applied to a 3D density structure, we use the total mass $M_5$ and average velocity $v_2$ of our simulated ejecta model.

The majority of the energy released after 0.1~days is in the form of $\gamma$-rays.
However, the ejecta trap only 60 percent of this energy at 0.1~days and progressively become optically thin to $\gamma$-rays, with the thermalization efficiency decreasing to about 10 percent at one day.
The form of \autoref{eq:fgamma} requires thermalization ratios to start from 1.0 at time $t=0$.

Since we assume that $\beta$ particles remain locally trapped, they do not diffuse out from the ejecta, and their thermalization efficiency decreases more gradually.
The efficiency of $\beta$ thermalization does decrease with the ongoing dilution of the expanding ejecta, causing emitted particles to deposit their kinetic energy on longer timescales \citep{Kasen:2019je}.
After a couple of hours, the primary contribution to energy deposition is from $\beta$ particles ($\alpha$ particles make a much smaller contribution).
The thermalization remains efficient out to timescales of days, only dropping below 90 percent at around two days.

We find that our thermalization ratios for $\alpha$ and $\beta$ particles are generally close to the analytical approximations over the first two days (less than about 5 percent difference) and agree fairly well (up to 20 percent difference) up to about five days.
Compared to the analytical formulae, which predict similar thermalization ratios of $\alpha$ and $\beta$ particles, we find that $\beta$ particles are more efficiently thermalized than $\alpha$ particles.
This is likely related to differences in the composition-dependent rates of individual $\alpha$ and $\beta^-$ decay reactions, which each emit particles with different energies and therefore have different thermalization timescales.

\bibliography{references_artis,references_papers}{}
\bibliographystyle{aasjournal}



\end{document}